\documentclass[aps,preprint]{revtex4}
\usepackage{amssymb,epsf}
\usepackage{amsmath,amsfonts}
\usepackage{graphicx}

\begin{document}

\title{Surface Terms of Quasitopological Gravity and Thermodynamics of Charged Rotating Black Branes}
\author{M. H. Dehghani$^{1,2}$\footnote{email address:
mhd@shirazu.ac.ir} and M. H. Vahidinia$^{1}$\footnote{email
address: vahidinia@shirazu.ac.ir}}

\affiliation{$^1$Physics Department and Biruni Observatory,
College of Sciences, Shiraz
University, Shiraz 71454, Iran\\
$^2$Research Institute for Astrophysics
and Astronomy of Maragha (RIAAM), Maragha, Iran\\
}

\begin{abstract}
We introduce the surface term for quasitopological gravity in order to
make the variational principle of the action well-defined. We also introduce
the charged black branes of quasitopological gravity and calculate the
finite action through the use of the counterterm method. Then we compute the thermodynamic quantities
of the black brane solution by use of Gibbs free energy and investigate the first law of thermodynamics by introducing
a Smarr-type formula. Finally, we generalize our solutions to the case of rotating charged solutions.
\end{abstract}
%\pacs{04.50.+h, 04.20.Jb, 04.70.Bw, 04.70.Dy}
\maketitle

\section{Introduction}

The AdS/CFT conjecture or its stronger version gauge/gravity
duality provides a new framework to holographic study of quantum
field theory in the nonperturbative regime. According to this
duality, the Einstein gravity in the bulk spacetime corresponds to
a gauge theory with a large number of colors $N_{c}$ and large 't
Hoof coupling $\lambda $ living on the boundary. Indeed in
principle, one can use the dictionary of duality and obtain
information on one side by performing computation on the other
side \cite {Maldacena:1997re}. Recently, it has been shown that
one can describe strongly coupling phenomena in QCD, condense
matter physics and superconductivity by performing gravity
calculations. For an interesting review on application in the
condense matter physics, see Ref. \cite {Hartnoll:2009sz}.

The gauge/gravity duality for Einstein gravity with a two-derivative bulk
action only works for large $N_{c}$ and large $\lambda $ limit. On the other
hand, the coupling constants in the gravity side relate to central charges
in the field theory. Thus, Einstein gravity does not have enough free
parameters to account for the ratios between central charges and therefore
it is only dual to those conformal field theories for which all the central
charges are equal. In order to extend the regime of validity to the case of
finite $\lambda $ and finite $N_{c}$, or field theories with different
ratios between central charges, one should add higher curvature terms or
higher derivative terms with more coupling constants within a perturbative
framework to action \cite{Buchel:2004di,Myers:2010jv}. The first-step
generalization is to add the Gauss-Bonnet term to the gravity action. But the
Gauss-Bonnet gravity has only one such a coupling constant so the range of
its dual theories is limited. In order to extend the duality to new classes
of field theories, one can consider terms with higher order curvature
interactions such as the third order Lovelock term \cite
{Hofman:2008ar,Myers:2010jv}. Although the equations of motion of third
order Lovelock gravity are second-order differential equations, the third order
Lovelock term has no contribution to the field equation in six or lower
dimensional spacetime and therefore it is useless for studying field
theories in four dimensions. Recently, a new toy model for gravitation action
has been proposed which includes cubic curvature interactions and has
contribution in five dimensions \cite{Myers:2010ru,Oliva:2010zd}. The
equations of motion of this theory are second-order differential equations
when the metric is spherically symmetry. As in the case of third order
Lovelock gravity, the cubic term vanishes in six dimensions. Although this
property of Lovelock gravity is due to a topological origin of Euler density
in six dimensions, this property of the new theory does not have a
topological origin \cite{Myers:2010ru}. Hence, this theory is known as
quasitopological gravity. The black hole solutions and holography study of
this toy model have been investigated in Refs. \cite{Myers:2010jv,QTG}.

It is known that the Einstein-Hilbert action does not have a
well-defined variational principle, since one encounters a total
derivative that produces a surface integral involving the
derivative of the variation of the metric normal to the boundary.
To cancel this normal derivative terms one has to add a surface
term to action \cite{Gib}. Also in the framework of the path
integral formalism of quantum gravity, the necessity of adding the
surface term has been emphasized \cite{Hawking:1979ig}.
Furthermore, surface terms should be known when one likes to study
the general relativity in the Hamiltonian framework \cite{York72}.
These reasons provide a strong motivation for introducing the
surface terms in higher curvature gravity such as Lovelock gravity
or quasitopological gravity. The suitable surface term for
Lovelock gravity in all orders has been introduced in \cite{MDDM}.
Here, the main aim of this paper is to introduce surface terms for
quasitopological gravity in order to have a well-defined
variational principle in the case of spacetimes with flat
boundary. We also present the static and rotating charged black
branes of quasitopological gravity and investigate their
thermodynamics. To obtain the conserved quantities of the
solutions, we use the counterterm method. This is due to the fact
that, as in the case of Einstein gravity, the action and conserved
quantities diverge when the boundary goes to infinity. We will
introduce a counterterm to deal with these divergences for
spacetimes with flat boundary. Since all curvature invariants of
the boundary are zero, the only possible boundary counterterm is
one proportional to the volume of the boundary regardless of the
number of dimensions \cite{counterterm}.

The outline of our paper is as follows. In Sec. \ref{Bound}, we give a brief
review of the quasitopological gravity and review the surface terms of the
Einstein and Gauss-Bonnet gravities. Then we introduce the surface term of
quasitopological gravity for spacetimes with flat boundary. In Sec. \ref
{Therm}, we introduce asymptotically AdS charged black branes of
quasitopological gravity and investigate the thermodynamic properties of
these solutions by using the relation between on shell action and Gibbs free
energy. Section \ref{rot} is devoted to the generalization of our static
charged solutions to the case of rotating charged solutions. We also study
the thermodynamic properties of charged rotating black branes. Finally, we
finish our paper with some closing remarks.

\section{Quasitopological Gravity and Surface Term \label{Bound}}

The action of quasitopological gravity in $(n+1)$ dimensions with negative
cosmological constant $\Lambda =-n(n-1)/2l^{2}$ in the presence of
the electromagnetic field may be written as
\begin{eqnarray}
I_{G} &=&\frac{1}{16\pi }\int_{\mathcal{M}}d^{n+1}x\sqrt{-g}\Big\{\frac{%
n(n-1)}{l^{2}}+R+\frac{\lambda l^{2}}{(n-2)(n-3)}\mathcal{L}_{2}  \notag \\
&&\hspace{1.5cm}+\frac{8(2n-1)\mu l^{4}}{(n-2)(n-5)(3n^{2}-9n+4)}\mathcal{X}%
_{3}-\frac{1}{4}F_{\mu \nu }F^{\mu \nu }\Big\},  \label{ActD}
\end{eqnarray}
where $F_{\mu \nu }=\partial _{\mu }A_{\nu }-\partial _{\nu }A_{\mu }$ is
the electromagnetic tensor field and $A_{\mu }$ is the vector potential. The
first and second terms are the well-known cosmological constant and
the Einstein-Hilbert action, the third term is the Gauss-Bonnet term

\begin{equation}
\mathcal{L}_{2}=R_{\mu \nu \gamma \delta }R^{\mu \nu \gamma \delta }-4R_{\mu
\nu }R^{\mu \nu }+R^{2},  \label{L2}
\end{equation}
and the fourth term is the quasitopological term \cite{Myers:2010ru}

\begin{eqnarray}
\mathcal{X}_{3} &=&R_{\mu \,\,\nu }^{\,\,\gamma \,\,\,\delta }R_{\gamma
\,\,\delta }^{\,\,\sigma \,\,\,\kappa }R_{\sigma \,\,\kappa }^{\,\,\mu
\,\,\,\nu }+\frac{1}{(2n-1)(n-3)}\Big\{\frac{3(3n-5)}{8}R_{\mu \nu \gamma
\delta }R^{\mu \nu \gamma \delta }R  \notag \\
&&-3(n-1)R_{\mu \nu \gamma \delta }R^{\mu \nu \gamma }{}_{\sigma }R^{\delta
\sigma }+3(n+1)R_{\mu \nu \gamma \delta }R^{\mu \gamma }R^{\nu \delta }
\notag \\
&& +6(n-1)R_{\mu \nu} R^{\nu \gamma }R_{\gamma }{}^{\mu } -\frac{3(3n-1)}{2}%
RR_{\mu \nu }R^{\mu \nu }+\frac{3(n+1)}{8}R^{3}\Big\}.
\end{eqnarray}
One should note that our coupling constant $\mu $ is $-\mu $ of Ref. \cite
{Myers:2010ru}. Indeed, this term has no contribution to field equations in
six or less than five-dimensional spacetimes. But, in contrast to the
Lagrangian of third order Lovelock gravity \cite{Lov} which has no
contribution to the field equations in five dimensions, this term
contributes to the field equation in five dimensions. The field equations of
quasitopological gravity are second-order differential equations only for
the case of static metric.

Gibbons and Hawking have mentioned that the Einstein-Hilbert action (with $%
\lambda =\mu =0$) does not have a well-defined variational principle, since
one encounters a total derivative which produces a surface integral
involving the derivative of variation of metric normal to the boundary, $%
\partial _r \delta g_{\mu \nu }$. Such terms, which make the variation of
action ill defined, should cancel by some surface terms which only depend on
the geometry of boundary. In fact, the normal derivative terms can be
canceled by the variation of Gibbons-Hawking surface term \cite{Gib}
\begin{equation}
I_{b}^{(1)}=\frac{1}{8\pi }\int_{\partial \mathcal{M}}d^{n}x\sqrt{-\gamma }K
\label{Ib1}
\end{equation}
where $\gamma _{ab}$ is induced metric on the boundary $\partial \mathcal{M}$
and $K$ is the trace of extrinsic curvature of the boundary. Therefore, any
generalization of Einstein-Hilbert action needs similar but much more
complicated surface terms. However, the surface terms which make the
variation of action of Lovelock gravity well-defined are known \cite{MDDM}.
This surface term for the case of Gauss-Bonnet gravity -with $\mu =0$ in the
action (\ref{ActD})- is $I_{b}^{(1)}+I_{b}^{(2)}$, where $I_{b}^{(2)}$ is
\cite{MDDM}
\begin{equation}
I_{b}^{(2)}=\frac{1}{8\pi }\int_{\partial \mathcal{M}}d^{n}x\sqrt{-\gamma }%
\left\{ \frac{2\lambda l^{2}}{(n-2)(n-3)}\left( J-2\hat{G}%
_{ab}^{(1)}K^{ab}\right) \right\} .  \label{Ib2}
\end{equation}
In Eq. (\ref{Ib2}) $\hat{G}_{ab}^{(1)}$ is the $n$-dimensional Einstein
tensor of the metric $\gamma _{ab}$, and $J$ is the trace of
\begin{equation}
J_{ab}=\frac{1}{3}%
(2KK_{ac}K_{b}^{c}+K_{cd}K^{cd}K_{ab}-2K_{ac}K^{cd}K_{db}-K^{2}K_{ab}).
\label{Jab}
\end{equation}

Now, we want to present the surface terms which make the variation of the
action of quasitopological gravity (\ref{ActD}) well defined. By using the
similarity of quasitopological gravity and third order Lovelock gravity, we
introduce a surface term which makes\ the action (\ref{ActD}) well defined.
For simplification, we restrict our consideration to the case of spacetimes
with flat boundary, i.e. $\widehat{R}_{abcd}(\gamma )=0$. Therefore, we
suggest a surface term with a combination of fifth order terms in extrinsic
curvature as
\begin{eqnarray}
&&I_{b}^{(3)}=\frac{1}{8\pi }\int_{\partial \mathcal{M}}d^{n}x\sqrt{-\gamma
}\Big\{\frac{3\mu l^{4}}{5n(n-2)(n-1)^{2}(n-5)}(\alpha _{1}K^{5}  \notag \\
&&\,\ \ \ \ \ \ \ \ \ \ \ \ \ \ \ \ \ \ \ \ \ \ \ \ \ \ \ \ +\alpha
_{2}K^{3}K_{ab}K^{ab}+\alpha
_{3}K_{ab}K^{ab}K_{d}^{c}K_{e}^{d}K_{c}^{e}+\alpha
_{4}KK_{b}^{a}K_{a}^{b}K_{d}^{c}K_{c}^{d}  \notag \\
&&\,\ \ \ \ \ \ \ \ \ \ \ \ \ \ \ \ \ \ \ \ \ \ \ \ \ \ \ \ +\alpha
_{5}KK_{b}^{a}K_{c}^{b}K_{d}^{c}K_{a}^{d}+\alpha
_{6}K_{b}^{a}K_{c}^{b}K_{d}^{c}K_{e}^{d}K_{a}^{e}+\alpha
_{7}K^{2}K_{b}^{a}K_{c}^{b}K_{a}^{c})\Big\}  \label{Ib3}
\end{eqnarray}
We can fix the unknown coefficients $\alpha _{i}$ such that all normal
derivative of\ $\delta g_{\mu \nu }$ of quasitopological term are canceled
by the variation of $I_{b}^{(3)}$. Since the quasitopological gravity gives
second-order differential equations only for the case of static solutions,
we present $I_{b}^{(3)}$ for the static metric. The $(n+1)$-dimensional static
metric with a flat boundary may be written as
\begin{equation}
ds^{2}=-g(r)dt^{2}+\frac{dr^{2}}{f(r)}+r^{2}\;\sum_{i=1}^{n-1}d\phi
_{i}{}^{2}.  \label{met1}
\end{equation}
Substituting the metric (\ref{met1}) in the action (\ref{ActD}) and
integrating by parts, we can choose $\alpha _{i}$'s such that the variation
of action becomes well defined. One encounters with only five equations for
the seven unknown coefficients, $\alpha _{i}$'s. So, one can set two of them
to zero, which we choose them to be $\alpha _{6}=\alpha _{7}=0$. It is worth
noting that the value of total action is independent of our choice. The
values of $\alpha _{i}$'s may be obtained as
\begin{eqnarray}
&&\alpha _{{1}}=n,\text{ \ \ }\alpha _{{2}}={-2,}\text{ \ \ }\alpha _{{3}%
}=4(n-1),  \notag \\
&&\alpha _{{4}}={-n(5n-6)},\text{ \ \ \ }\alpha _{{5}}={(n-1)(5n-6)}.
\end{eqnarray}

In general $I=I_{G}+I_{b}^{(1)}+I_{b}^{(2)}+I_{b}^{(3)}$ is divergent when
evaluated on solutions. One way of eliminating such divergence terms is
through the use of the background subtraction method \cite{subtrac}. However for
asymptotically AdS solutions, one can instead deal with these divergences by
using the counterterm method inspired by AdS/CFT correspondence. In this
method the divergence terms are removed by adding a counterterm action $I_{%
\mathrm{ct}}$ which depends on the boundary curvature invariants. The
suitable counterterm for quasitopological gravity is unknown. However, for
the case of spacetimes with flat boundary $\hat{R}_{abcd}(\gamma )=0$, one
can add a term proportional to the volume of the boundary:
\begin{equation}
I_{\mathrm{ct}}=-\frac{1}{8\pi }\int_{\partial \mathcal{M}}d^{n}x\sqrt{%
-\gamma }\;\frac{(n-1)}{L_{\mathrm{eff}}},  \label{Ict}
\end{equation}
where $L_{\mathrm{eff}}$ is a scale length factor that depends on $l$, $\mu $,
and $\lambda $. Of course, $L_{\mathrm{eff}}$ must reduce to $l$ as $\mu $
and $\lambda $ go to zero.

\section{Thermodynamics of Charged black Branes\label{Therm}}

Here, we investigate the thermodynamics of static charged solutions of
quasitopological gravity by the counterterm method introduced in the last
section. Since we are assuming spherical symmetry, we can obtain the
one-dimensional action. Using the static metric (\ref{met1}) with $%
g(r)=N^{2}(r)f(r)$ and
\begin{equation}
A_{\mu }=h(r)\delta _{\mu }^{0}
\end{equation}
for the vector potential, after integration by parts, the one-dimensional
action may be written as
\begin{equation}
I_{G}=\frac{1}{16\pi }\int_{\mathcal{M}}d^{n+1}x\left\{
\frac{n-1}{l^{2}}N(r)[r^{n-6}(-\mu l^{6}f^{3}+\lambda
l^{4}r^{2}f^{2}-r^{4}l^{2}f+r^{6})]^{\prime }+\frac{r^{n-1}}{2N(r)}%
h^{\prime 2}\right\}   \label{Act2}
\end{equation}
where prime denotes the derivative with respect to $r$. Varying the action
with respect to $f(r)$ yields
\begin{equation}
\left( 1-2\lambda \frac{l^{2}}{r^{2}}f(r)+3\mu \frac{l^{4}}{r^{4}}%
f^{2}(r)\right) \frac{dN(r)}{dr}=0,  \label{eom1}
\end{equation}
which shows that $N(r)$ should be a constant which we set equal to $1$.
Varying the action with respect to $h(r)$ and substituting $N(r)=1$ gives
\begin{equation}
rh^{\prime \prime }+(n-1)h^{\prime }=0,  \label{eom2}
\end{equation}
which has the solution
\begin{equation}
h(r)=-\sqrt{\frac{2(n-1)}{n-2}}\frac{q}{r^{n-2}}.  \label{hr}
\end{equation}
After varying the action (\ref{Act2}) with respect to $N(r)$, substituting $%
N(r)=1$ and using Eq. (\ref{hr}), one obtains
\begin{equation}
1-\Psi +\lambda \Psi ^{2}-\mu \Psi ^{3}=\frac{ml^{2}}{r^{n}}-\frac{%
q^{2}l^{2}}{r^{2(n-1)}},  \label{Eqsi}
\end{equation}
where $\Psi (r)=l^{2}r^{-2}f(r)$ and $m$ is an integration constant. The
three roots of cubic Eq. (\ref{Eqsi}) are
\begin{eqnarray}
f_{1}(r) &=&\frac{r^{2}}{l^{2}}\left( \frac{\lambda }{3\mu }+\alpha +{u}\,{%
\alpha ^{-1}}\right) ,  \label{f1} \\
f_{2,3}(r) &=&\frac{r^{2}}{l^{2}}\left\{ \frac{\lambda }{3\mu }-\frac{1}{2}%
(\alpha +u\,{\alpha ^{-1}})\pm \frac{\sqrt{3}}{2}(\alpha -{u}\,{\alpha ^{-1}%
})\right\} ,  \label{f2}
\end{eqnarray}
where
\begin{eqnarray}
u &=&\frac{\lambda ^{2}-3\mu }{9\mu ^{2}},\text{ \ }\ \alpha =\left( v+\sqrt{%
v^{2}-u^{3}}\right) ^{1/3},\  \\
v &=&\frac{2\lambda ^{3}-9\mu \lambda }{54\mu ^{3}}+\frac{1}{2\mu }(1-\frac{%
ml^{2}}{r^{n}}+\frac{q^{2}l^{2}}{r^{2(n-1)}}).
\end{eqnarray}
All of the three roots could be real in the appropriate range of $\mu $ and $%
\lambda $ \cite{Myers:2010ru}. For example one can rewrite the last two
solutions as
\begin{equation}
f_{2,3}(r)=\frac{r^{2}}{l^{2}}\left\{ \frac{\lambda }{3\mu }+\sqrt{u}\left(
\cos \frac{\theta }{3}\pm \sqrt{3}\sin \frac{\theta }{3}\right) \right\} ,
\end{equation}
where $\cos {\theta }=vu^{-3/2}$ and $\sin {\theta }=\sqrt{u^{3}-v^{2}}%
u^{-3/2}$. The second and third solutions are real provided $u^{3}>v^{2}$.
In the following, we will consider only the first solution $f_{1}(r)$, which
is real every where provided $u^{3}<v^{2}$. This condition reduces to
\begin{equation}
\mu >\frac{\lambda }{3}-\frac{2}{27}\left\{ 1-\left( 1-3\lambda \right)
^{3/2}\right\} ,
\end{equation}
which can be satisfied provided $\lambda \leq 1/3$.

\begin{figure}[tbp]
\centering {\includegraphics[width=7cm]{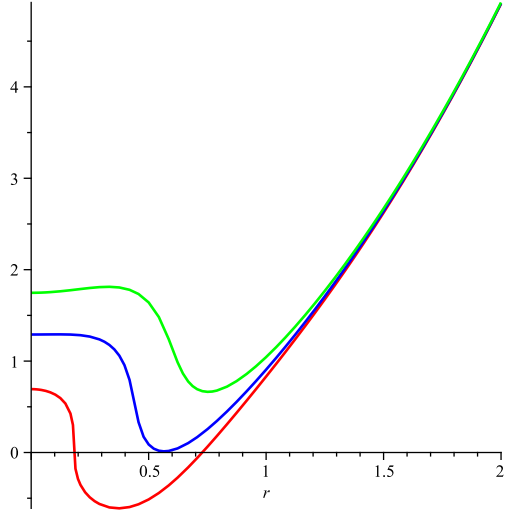}}
\caption{$f(r)$ vs $r$ for $n=4$, $l=1$, $\protect\lambda =0.2$, $\protect%
\mu =0.03$, $m=0.3$, $q>q_{\mathrm{ext}}$, $q=q_{\mathrm{ext}}$ and $q<q_{\mathrm{ext}}
$ from up to down, respectively.}
\label{Fr}
\end{figure}

One can obtain the temperature of the event horizon by the standard method of
analytic continuation of the metric as
\begin{equation}
T=\beta ^{-1}=\frac{f^{\prime }(r_{+})}{4\pi }=\frac{%
n-(n-2)q^{2}l^{2}r_{+}^{2-2n}}{4\pi l^{2}}r_{+},  \label{T1}
\end{equation}
where $r_{+}$ is the outer horizon radius which is the largest real root of $%
r_{+}^{2n-2}-ml^{2}\ r_{+}^{n-2}+q^{2}l^{2}=0.$ The solution given
in Eq. (\ref{f1}) presents a black hole with two horizons provided
$q<q_{\mathrm{ext}}$, an
extreme black hole for $q=q_{\mathrm{ext}}$ and a naked singularity if $q>q_{%
\mathrm{ext}}$, where (see Fig. \ref{Fr})
\begin{equation*}
q_{\mathrm{ext}}=\sqrt{\frac{n}{n-2}}\frac{r_{\mathrm{ext}}^{n-1}}{l}.
\end{equation*}

The electric potential $\Phi $, measured at infinity with respect to the
horizon, is defined by \cite{Gub}
\begin{equation}
\Phi =A_{\mu }\xi ^{\mu }\mid _{r\rightarrow \infty }-A_{\mu }\xi ^{\mu
}\mid _{r=r_{+}}, \label{Pot}
\end{equation}
where $\xi ^{\mu }$ is the null generator of the
horizon. For our static case, $\xi ^{\mu }=\delta _{t}^{\mu }$ and therefore one obtains
\begin{equation}
\Phi =\sqrt{\frac{2(n-1)}{n-2}}\frac{q}{r_{+}^{n-2}}.  \label{Phi1}
\end{equation}

We calculate entropy, electric charge and mass of this black hole through
the use of Gibbs free energy. Following \cite{Gib}, one can identify the
Gibbs free energy function $G$ with the Euclidean action times the
temperature, $G=TI_{\mathrm{E}}$. The finite total action, $I_{\mathrm{E}%
}=I_{\mathrm{g}}+I_{\mathrm{b}}+I_{\mathrm{ct}}$ may be found through the
use of the counterterm method introduced in the last section. Straightforward
calculations shows that the total Euclidian action is finite provided one
sets $L_{\mathrm{eff}}$ in Eq. (\ref{Ict}) equal to

\begin{equation}
L_{\mathrm{eff}}=\frac{30\Psi _{1\infty }^{1/2}}{3\mu \Psi
_{1\infty }^{3}-5\lambda \Psi _{1\infty }^{2}+15\Psi _{1\infty
}+15}l,
\end{equation}
where $\Psi _{1\infty }=\lim_{r\rightarrow \infty }\Psi _{1}(r).$ The finite
action per volume is calculated as
\begin{equation}
I_{\mathrm{E}}=-\frac{\beta }{16\pi l^{2}}(r_{+}^{n}+\frac{q^{2}l^{2}}{%
r_{+}^{n-2}}).  \label{FAction}
\end{equation}
Now using Eqs. (\ref{T1}) and (\ref{Phi1}), the Gibbs free energy per unit
volume in terms of the intensive quantities $T$ and $\Phi $ may be written
as
\begin{equation}
G(T,\Phi )=-\frac{1}{16\pi l^{2}}\left( r_{+}^{n}+\frac{n-2}{2(n-1)}%
l^{2}\Phi ^{2}r_{+}^{n-2}\right) ,\,
\end{equation}
where $r_{+}$ in terms of temperature and electric potential is
\begin{equation}
\ \ \ \ \ r_{+}=\frac{2\pi l^{2}}{n}T+2\left[ \frac{\pi ^{2}l^{4}}{n^{2}}%
T^{2}+\frac{(n-2)^{2}l^{2}}{8n(n-1)}\Phi ^{2}\right] ^{1/2}.
\end{equation}
The \ entropy, charge, and energy per volume of charged black brane may be
calculated as
\begin{eqnarray}
S &=&-\left( \frac{\partial G}{\partial T}\right) _{\Phi }=\frac{1}{4}%
r_{+}^{n-1},  \label{Ent1} \\
Q &=&-\left( \frac{\partial G}{\partial \Phi }\right) _{T}=\frac{\sqrt{%
2(n-1)(n-2)}}{16\pi }q,  \label{Ch1} \\
M &=&G+TS+\Phi Q=\frac{n-1}{16\pi }m
\end{eqnarray}
One may note that the entropy density obtained in Eq. (\ref{Ent1}) is in
agreement with the entropy density calculated in Ref. \cite{Myers:2010ru}.
Also, the charge density of Eq. (\ref{Ch1}) is the same as the value that
can be calculated by use of the Gauss law.

One can also find a Smarr-type formula for energy density as a function of $S
$ and $Q$:
\begin{equation}
M(S,Q)=\frac{n-1}{\pi l^{2}}\left[ S^{2}+\frac{8\pi ^{2}l^{2}}{(n-1)(n-2)}Q^{2}%
\right] (4S)^{(2-n)/(n-1)}.
\end{equation}
Now one may regard the parameters $S$ and $Q$ as a complete set of extensive
parameters for the energy density $M(S,Q)$ and define the intensive
parameters conjugate to $S$ and $Q$. These quantities are the temperature
and the electric potential:
\begin{equation}
T=\left( \frac{\partial M}{\partial S}\right) _{Q},\,\ \ \ \ \Phi =\left(
\frac{\partial M}{\partial Q}\right) _{S}.  \label{Inten}
\end{equation}
It is a matter of straightforward calculation to show that the intensive
quantities calculated by Eq. (\ref{Inten}) coincide with Eqs. (\ref{T1}) and
(\ref{Phi1}) found previously. Thus, the thermodynamic quantities calculated
in this section satisfy the first law of thermodynamics:
\begin{equation}
dM=TdS+\Phi dQ.
\end{equation}

\section{$(n+1)$-dimensional Charged Rotating Black Branes \label{rot}}

In this section we like to endow our charged static solution with a global
rotation. The rotation group in $n$ dimensions is $SO(n)$ and therefore the
number of independent rotation parameters is $[n/2]$, where $[x]$ is the
integer part of $x$. The metric of a $D$-dimensional asymptotically AdS
rotating solution with $k\leq \lbrack n/2]$ rotation parameters whose
constant $(t,r)$ hypersurface has zero curvature may be written as \cite
{Awad:2002cz}
\begin{eqnarray}
ds^{2} &=&-N(r)^{2}f(r)\left( \Xi dt-{{\sum_{i=1}^{k}}}a_{i}d\phi
_{i}\right) ^{2}+\frac{r^{2}}{l^{4}}{{\sum_{i=1}^{k}}}\left( a_{i}dt-\Xi
l^{2}d\phi _{i}\right) ^{2}  \notag \\
&&\ +\frac{dr^{2}}{f(r)}-\frac{r^{2}}{l^{2}}{\sum_{i<j}^{k}}(a_{i}d\phi
_{j}-a_{j}d\phi _{i})^{2}+r^{2}\;\sum_{i=k+1}^{n-1}d\phi _{i}{}^{2},  \notag
\\
\Xi ^{2} &=&1+\sum_{i=1}^{k}\frac{a_{i}^{2}}{l^{2}},  \label{met2}
\end{eqnarray}
where the angular coordinates are in the range $0\leq \phi _{i}<2\pi $. The
periodic nature of $\phi _{i}$ allows the metrics (\ref{met1}) and (\ref
{met2}) to be locally mapped into each other but not globally, and so they
are distinct \cite{Stach}. Also the gauge potential for this solution is
given by
\begin{equation}
A_{\mu }=-\sqrt{\frac{2(n-1)}{n-2}}\frac{q}{r^{n-2}}\left( \Xi
dt-\sum_{i=1}^{k}a_{i}d\phi _{i}\right) .
\end{equation}

Now, we investigate the thermodynamics of rotating charged black branes. The
temperature of the event horizon is given by
\begin{equation}
T=\frac{1}{2\pi }\left( -\frac{1}{2}\nabla _{b}\xi _{a}\nabla ^{b}\xi
^{a}\right) _{r=r_{+}}^{1/2},  \label{Tem1}
\end{equation}
where $\xi $ is the Killing vector
\begin{equation}
\xi =\partial _{t}+\sum_{i=1}^{k}\Omega _{i}\partial _{\phi _{i}} \label{xii}
\end{equation}
and $\Omega _{i}$ is the angular velocity of the Killing horizon given as
\begin{equation}
\Omega _{i}=-\left[ \frac{g_{t\phi _{i}}}{g_{\phi _{i}\phi _{i}}}\right]
_{r=r_{+}}=\frac{a_{i}}{\Xi l^{2}}.  \label{Om}
\end{equation}
Using Eqs. (\ref{Tem1})-(\ref{Om}), one obtains
\begin{equation}
T=\beta ^{-1}=\frac{f^{\prime }(r_{+})}{4\pi \Xi }=\frac{%
n-q^{2}l^{2}(n-2)r_{+}^{2-2n}}{4\pi l^{2}\Xi }r_{+}\,.  \label{Tem2}
\end{equation}

The electric potential $\Phi $ can be calculated by use of Eqs. (\ref{Pot}) and (\ref{xii}) as
\begin{equation}
\Phi =\sqrt{\frac{2(n-1)}{n-2}}\frac{q}{\Xi r_{+}^{n-2}}.
\label{Ph2}
\end{equation}

Again, one may obtain the finite action per volume through the use of the
counterterm method as
\begin{equation}
I_{\mathrm{E}}=-\frac{\beta }{16\pi l^{2}}(r_{+}^{n}+\frac{q^{2}l^{2}}{%
r_{+}^{n-2}}).
\end{equation}
Now using Eqs. (\ref{Om}), (\ref{Tem2}) and \ref{Ph2}), the Gibbs free
energy per unit volume in terms of the intensive quantities $T$, $\Phi $ and
$\Omega _{i}$'s may be written as
\begin{equation}
G(T,\Phi ,\mathbf{\Omega })=-\frac{1}{16\pi l^{2}}\left( r_{+}^{n}+\frac{%
(n-2)l^{2}\Phi ^{2}}{2(n-1)(1-l^{2}\mathbf{\Omega }^{2})}r_{+}^{n-2}\right) ,
\end{equation}
where the horizon radius in term of the intensive quantities $T$, $\Phi $
and $\Omega _{i}$'s is
\begin{equation}
r_{+}=(1-l^{2}\mathbf{\Omega }^{2})^{-1/2}\left\{ \frac{2\pi l^{2}}{n}T+2%
\left[ \frac{\pi ^{2}l^{4}}{n^{2}}T^{2}+\frac{(n-2)^{2}l^{2}}{8n(n-1)}\Phi
^{2}\right] ^{1/2}\right\} .
\end{equation}
Now\ entropy, charge and energy per volume of rotating charged black brane
may be calculated as
\begin{eqnarray}
S &=&-\left( \frac{\partial G}{\partial T}\right) _{\Phi ,\Omega _{i}}=\frac{%
\Xi }{4}r_{+}^{n-1}, \\
Q &=&-\left( \frac{\partial G}{\partial \Phi }\right) _{T,\Omega _{i}}=\frac{%
\sqrt{2(n-1)(n-2)}}{16\pi }\Xi q,\, \\
J_{i} &=&-\left( \frac{\partial G}{\partial \Omega _{i}}\right) _{T,\Phi }=%
\frac{n}{16\pi }\Xi ma_{i}, \\
M &=&G+TS+\Phi Q+\sum\limits_{i=1}^{k}\Omega _{i}J_{i}=\frac{1}{16\pi }(n\Xi
^{2}-1)m.
\end{eqnarray}

\section{Concluding Remarks \label{Conc}}

The action of quasitopological gravity does not have a
well defined variational principle, and therefore one needs to add
a surface term in order to cancel the derivative of the variation
of the metric normal to the boundary. Also, one needs to know the
surface terms in the framework of the path integral formalism of
quantum gravity, or if one likes to study the general relativity
in Hamiltonian framework. Motivated by these facts, we introduced
the surface term that makes the variation of the action
well-defined. We also introduced the counterterm to deal with the
divergences that appeared in the action and calculated the finite
action through the use of the counterterm method. Obtaining the finite
action, we presented the Gibbs free energy of static charged black
branes as a function of the intensive thermodynamic quantities,
temperature, and electric potential. We calculated the
thermodynamic quantities $S$, $Q$ and energy \ densities of the
solution through the use of Gibbs free energy and found that these
quantities coincide with their values calculated by the geometrical
method. We also showed that these quantities satisfy the first law
of thermodynamics by obtaining a Smarr-type formula for energy
density as a function of entropy and charge densities.

Next, we generalized the static charged solutions to rotating charged
solutions and calculated all the conserved and thermodynamic quantities of
these black brane solutions. One may use the boundary term introduced in
this paper to consider the thermodynamics of the Lifshitz black brane of
quasitopological gravity introduced in \cite{BDM}.

\acknowledgements
This work was supported by the Research Institute for Astrophysics and
Astronomy of Maragha. M. H. Vahidinia would like to thank S. Jalali and N.
Farhangkhah for useful discussion.

\end{document}